# Data Acquisition System for CSNS Neutron Beam Monitor

Jian ZHUANG[1,2,3], Ke ZHOU[2,3], Lijiang LIAO[2,3], Lei HU[2,3], Jiajie LI[2,3], Yongxiang QIU[2,3]

State Key Laboratory of Particle Detection and Electronics[1], Beijing, P.R.China
Institute of High Energy Physics[2], Beijing 100049, P.R.China
Dongguan Neutron Science Center[3], Dongguan 523803, P.R.China

*Abstract*–In Chinese Spallation Neutron Source (CSNS), proton beam is used to hit metal tungsten target, and then high flux neutrons are generated for experiments on instruments. For neutron flux spectrum correction, boron-coated GEM, lithium glass and $^3$He are used as neutron beam monitor in instruments.

To be integrated into neutron instrument control, a new DAQ software for neutron beam monitor is developed, called NEROS (Neutron Event Readout System). NEROS is based on EPICS V4 and a unified data format for CSNS neutron beam monitor is defined. The framework and software design of NEROS is introduced in this paper, including real-time data readout, data processing, data visualization and data storage in Nexus format. The performance is evaluated through offline test, X-ray test and $^{252}$Cf test. The deployment of NEROS and its running result in CSNS instrument commissioning are also introduced in this paper.

## I. INTRODUCTION

Neutron scattering is a powerful method to probe the structure of the microscopic world, becoming a complementary technique to x-ray in the advanced researches in physics, chemistry, biology, life science, material science, new energy, as well as in other applications. To meet the increasing demands from user community, China decided to build a world-class spallation neutron source, called CSNS. It can provide a neutron scattering platform with high flux, wide wavelength range and high efficiency to users. The pulsed-beam feature allows studies not only on the static structure but also the dynamic mechanisms of the microscopic world.

CSNS mainly consists of an H-Linac and a proton rapid cycling synchrotron [1]. It is designed to accelerate proton beam pulses to 1.6GeV kinetic energy at 25Hz repetition rate. Proton pulses strike a solid metal target to produce spallation neutrons. The facility of CSNS is shown in figure Fig.1.

**Jian** ZHUANG(1976.10- ), male, received his doctor's degree in computer applied technology from University of University of Chinese Academy of Sciences. He current works as associate professor at Institute of High Energy Physics in China. His research interest includes control system, real-time system and DAQ.( e-mail: zhuangj@ihep.ac.cn)

Corresponding author: Jian ZHUANG, (e-mail:zhuangj@ihep.ac.cn)

In CSNS, boron-coated GEM, lithium glass and $^3$He tube are used as neutron beam monitor. All the neutron beam monitor are list in table 1.

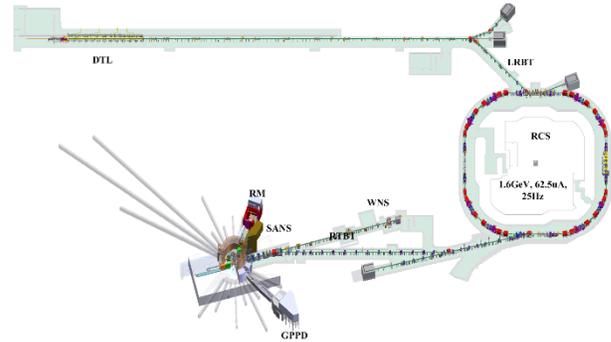

Fig.1. Overall construction of CSNS

The experimental control system of CSNS is in charge of target and instrument control [3][4]. The task of instrument control mainly includes providing local control to instrument device, integrating all devices belonged to the instrument into one system, providing facility information, mainly on accelerator information, providing the trigger signal (T0), and synchronizing all system. To simplify design of neutron instrument, the DAQ software of neutron beam monitor must to be integrated into instrument control. A new DAQ software based on EPICS V4 and aeraDetector is developed, called NEROS (Neutron Event Readout System) [2].

Table 1: Beam Monitor in CSNS

| Instrument | quantity | Type of neutron beam monitor |
|---|---|---|
| GPPD | 3 | M1: GEM |
|  |  | M2: GEM |
|  |  | M3: GEM |
| MR | 1 | M1: GEM |
| SANS | 3 | M1: GEM |
|  |  | M1: $^3$He |
|  |  | M1: Lithium glass |

In this paper, the framework design of NEROS is introduced, including real-time data readout, data processing, data visualization and data storage in Nexus format. The performance is tested and discussed by offline test, X-ray test and $^{252}$Cf test. The deployment of NEROS and its running result in CSNS instrument commissioning are also introduced in this paper.

## II. DETECTORS AND ELECTRONICS

The neutron beam monitor consists of detector, electronics, network and readout computer as shown in Fig2. For widly used GEM detector, the sensitive area is 100mm × 100mm, and the neutron detection efficiency is about 1‰. The electronics is based on Xilinx Spartan 6 FPGA.

The signal from detector is converted into digital through front amplifier, and transferred into FPGA. The hit position is determined by all adjacent digital signal through barycenter method. All hit information are packaged in a specific format and marked with high precision timestamp, then send to readout computer by Ethernet [5][6][7]. The TCP protocol is used in data transfer and UDP protocol is used for control and configuration.

## III. FRAMEWORK OF NEROS

NEROS software consists of epicsROS, visualization, data storage and other client, as shown in Fig.2.

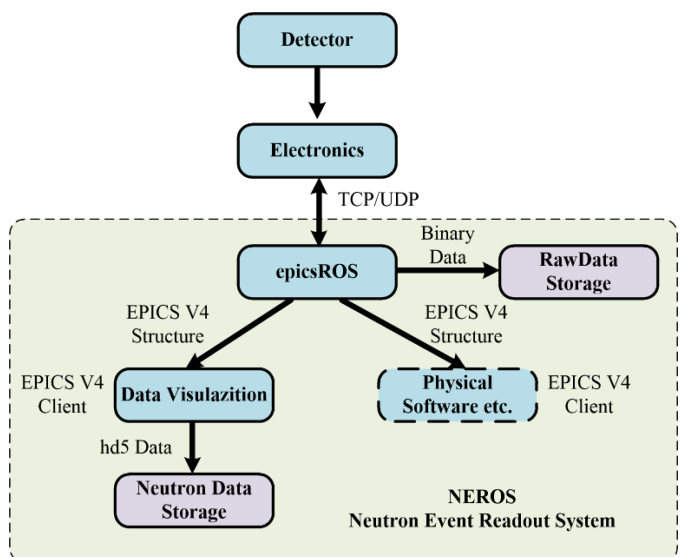

Fig.2. Framework of NEROS

After neutron data from electronics received by TCP protocol, the epicsROS will match the beginning and ending pair of package to check out the whole data of proton pulse. Then, the raw data can be stored on disk array for machine diagnosis.

The neutron hit information are processed in NEROS software. The X and Y of hit position are converted into ID of neutron detector pixel. The neutron TOF (time of flight) is also calculated and corrected in NEROS. All these information will be packaged into unified format for all types of neutron beam monitor, as Fig.3 shown. Proton pulse information, such as time of proton hit target(T0), proton pulse id are also included in this package. The unified package format of different neutron detector can simplify data processing and data storage.

After data processed in epicsROS, the neutron package will send to EPICS V4 client pulse by pulse. The epicsROS can support multiple clients. Data visualization software and online analysis software can be registered in epicsROS as EPICS V4 client.

```
neutrons
structure
    time_t timeStamp 2018-1-25T08:19:12.634 18
    epics:nt/NTscalar:1.0 proton_Charge
        double value 1.03+12
    epics:nt/NTscalar:1.0 sequence
        uint value 92347
    epics:nt/NTScalarArray:1.0 time_of_flight
        unit[] value[728, 733, 745, ··· 767,794,823]
    epics:nt/NTScalarArray:1.0 pixel_id
        unit[] value[1284, 932, 458, ··· , 78, 532, 987]
```

Fig.3. Unified Neutron Data Structure

## IV. READOUT SOFTWARE

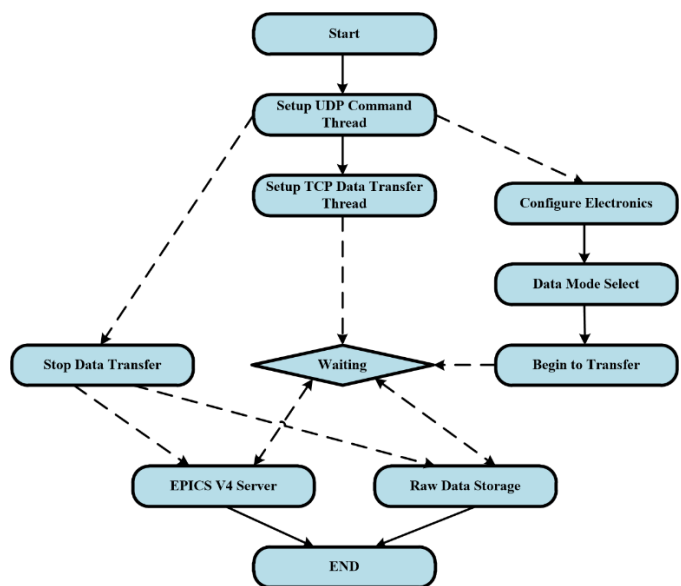

Fig. 4 Workflow of epicsROS

In NEROS, epicsROS is in charge of configure electronics and readout the raw neutron data from FPGA. After the epicsROS is started，command thread and data thread are setup, waiting the control command from operator interface. Before taking data, electronics need to be configured by UDP command. The threshold of each electronics channel, running mode and so on are set. When the START command from PV received, epicsROS set electronics to start running. The neutron data received from electronics are stored as raw data in readout computer, then is processed in NEROS and broadcasted to EPICS V4 client. This flow is shown in Fig.4.

In order to use the CPU of readout computer efficiently, there are three thread in epicsROS and synchronized by interlocking signals, including receiving data, analysis data, and distribution data, as shown in Fig. 5. To isolate these threads, there are two ring-buffers between threads. After neutron data from electronics received, data receiving thread stores data in receiving ring-buffer and set signal to inform analysis thread starting data analysis. When data analysis is done or incomplete data is met, the analysis thread is paused until new data is arrived. The data already processed are stored in sending ring-buffer. The distribution thread broadcasts these data to EPICS V4 client in uniform data format. The receiving thread is set the highest priority to ensure neutron data can be received in time.

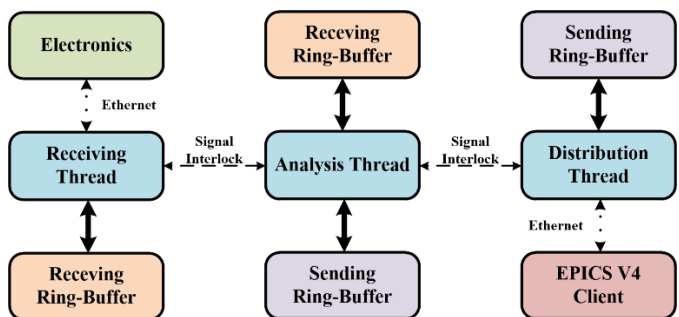

Fig.5 Threads and Ring-Buffer of epicsROS

## V. Data Visualization and Storage

Data visualization and storage software are based on EPICS V4 client. After connected to epicsROS, monitor event module will always listen neutron data broadcast of EPICS V4 server in epicsROS. When the neutron data arrived, EventHandler function will store the neutron data into NDArray according to the detector ID. The plugins are used to real-time processing and statistical analysis of neutron data in NDArray. These functions are mainly for real-time analysis, such as position distribution and time distribution of neutron TOF. Then these result are showed on the operator interface online.

NEROS uses Nexus file format to store neutron data for offline use. Nexus file can hold the information of the experimental configuration, the description of the instrument, the sample and the sample environment information, the data of the neutron experiment and all the other information that can help to understand the experiment on the neutron instrument in a formatted way.

For one neutron experiment, NEROS provides stream mode and capture mode for neutron data storage. A single run is saved into one file in stream mode. For capture mode, the neutron data are saved as multiple files, according to the run time.

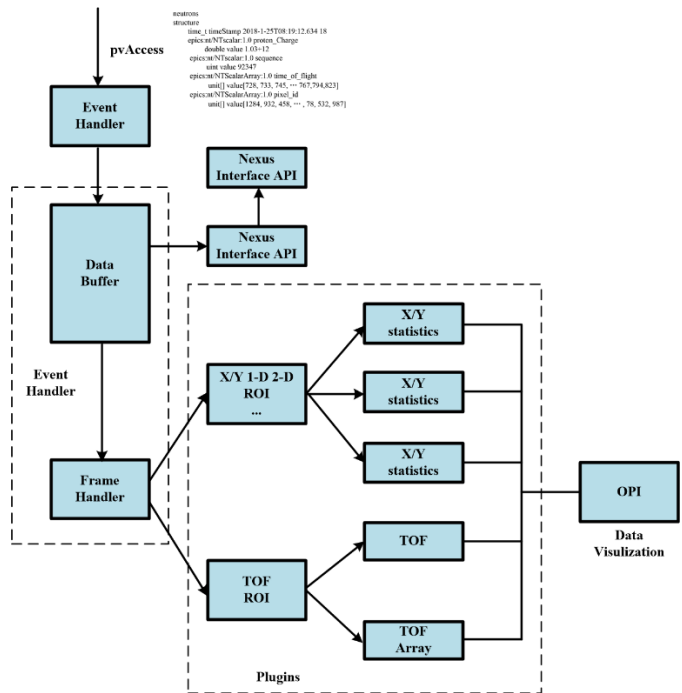

Fig.5 Data Flow of Visualization and Storage

Metadata of experiment and the neutron events as a data group can be stored into one Nexus file. Now, in CSNS, the metadata of detector including high voltage and gas are stored in Nexus file, as shown in Fig. 6. The dataset consists of the number of proton pulse, pixel id and time of flight in this pulse.

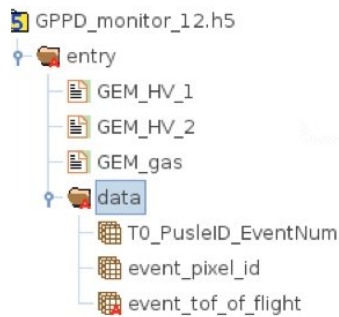

Fig. 6 Structure of Nexus file for neutron monitor

## VI. Performance Evaluation

To evaluate the performance of NEROS, the data throughput and CPU usage on Intel Xeon CPU E5-2609 v3 @ 1.90GHz are tested. A software simulator are developed to simulate electronics. The number of neutron event of a proton pulse can be set in simulator. And a data inspector in EPICS client can check the validation of data. The CPU usage is monitored while event number of one pulse is increased. The Ethernet throughput is calculated out using data format.

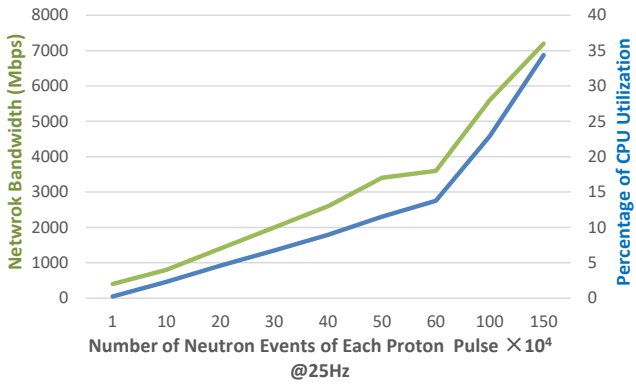

Fig.7 the Performance of NEROS Framework

The evaluation result of NEROS framework is shown in Fig.7. The neutron event can achieve $1.5\times10^6$ neutron events in 1 proton pulse @25Hz while the CPU usage is about 35%. And the Ethernet throughput is about 7Gbps in 10Gbps net. The rest of ability CPU can be used to neutron data process.

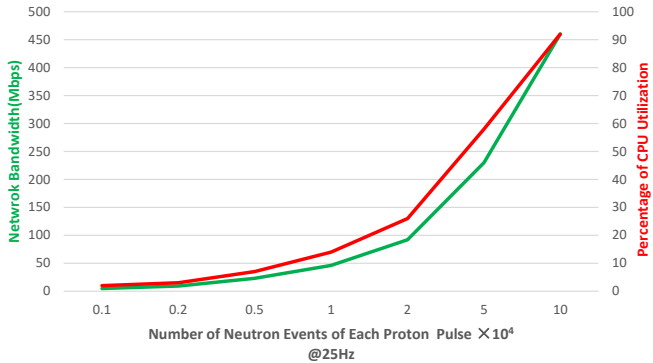

Fig.8 the Performance of NEROS Framework

For the NEROS with real neutron data process, the performance is also evaluated, as shown in Fig. 8. Because of the high CPU utilization of data analysis thread, the NEROS only can achieve $1\times10^5$ neutrons events in 1 proton pulse @25Hz. That means the NEROS can work in $2.5\times10^6$ neutron event per second. According to physical estimation, there are $2.5\times10^5$ neutron events per second on neutron beam monitor. The capacity of NEROS is enough for neutron monitor in CSNS instrument. In the future, the thread of analysis will be split into several parallelized threads to improve NEROS performance.

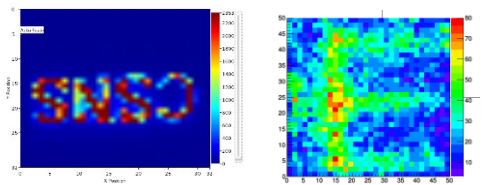

(a) "CSNS" mask under x-ray    (b) "E" under $^{252}$Cf

Fig.7. Mask Test under x-ray and neutron source

## VII. RUNNING ON CSNS INSTRUMENT

From Nov.1 2017 to Feb. 28 2018, CSNS has completed the joint commissioning of accelerator, target and neutron instruments. In this period, 5.4MWhr proton power was accumulated on the target, and the 9 neutron monitors, including detectors, electronics and NEROS software runs steadily, as shown in Fig.8. According to cross check of other detector, the histogram and processed data are reliability.

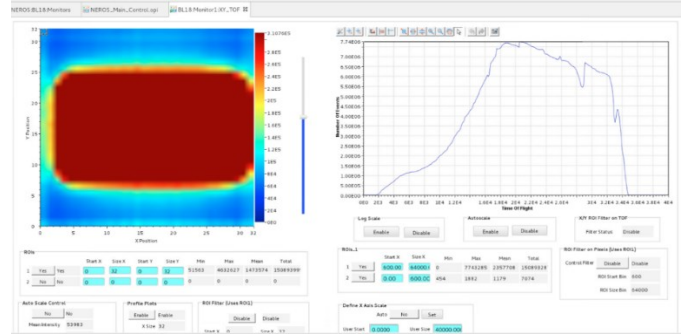

Fig.8. NEROS in Real Neutron Experiment

## VIII. CONCLUSION

In the Chinese Spallation Neutron Source, the NEROS (Neutron Event Readout System) software is developed for data acquisition of neutron beam monitor. The performance is evaluated under throughout test and mask test. After commission in real neutron experiment, the NEROS can achieve the request of CSNS. The parallelization of analysis threads will be carefully tuning to performance improvement in the future.

## IX. ACKNOWLEDGEMENTS

This work was supported by China Spallation Neu-tron Source. This work is also was support by the science and technology project of Guangdong province under grand No. 2016B090918131 and 2017B090901007.